\documentclass[prc,preprint,showpacs,preprintnumbers,amsmath,amssymb]{revtex4}

\usepackage[mathscr]{eucal}
\usepackage{graphicx}
\def\slr#1{\setbox0=\hbox{$#1$}           
   \dimen0=\wd0                                 
   \setbox1=\hbox{/} \dimen1=\wd1               
   \ifdim\dimen0>\dimen1                        
      \rlap{\hbox to \dimen0{\hfil/\hfil}}      
      #1                                        
   \else                                        
      \rlap{\hbox to \dimen1{\hfil$#1$\hfil}}   
      /                                         
   \fi}

\def\kp{k^{\,\prime}}
\def\kpsq{k^{\,\prime\,2}}
\def\ksq{k^2}
\def\kdp{k^{\,\prime\prime}}
\def\myint#1{\!\int\!\!\frac{d^4\!{#1}}{(2\pi)^4}\,}
\def\gev#1{ GeV${}^{#1}$}
\def\be{\begin{eqnarray}}
\def\ee{\end{eqnarray}}

\renewcommand{\theequation}%
    {\arabic{section}.\arabic{equation}}
\makeatletter \@addtoreset{equation}{section} \makeatother

\begin{document}


\title{Quark Condensates and Momentum-Dependent Quark Masses
 in a Nonlocal Nambu--Jona-Lasinio Model}

\author{Bing He}
\author{Hu Li}
\author{Qing Sun}
\author{C.M. Shakin}
 \email[email:]{casbc@cunyvm.cuny.edu}
\affiliation{%
Department of Physics and Center for Nuclear Theory\\
Brooklyn College of the City University of New York\\
Brooklyn, New York 11210
}%

\date{January, 2002}

\begin{abstract}
The Nambu--Jona-Lasinio (NJL) model has been extensively studied
by many researchers. In previous work we have generalized the NJL
model to include a covariant model of confinement. In the present
work we consider further modification of the model so as to
reproduce the type of Euclidean-space momentum-dependent quark
mass values obtained in lattice simulations of QCD. This may be
done by introducing a nonlocal interaction, while preserving the
chiral symmetry of the Lagrangian. In other work on nonlocal
models, by other researchers, the momentum dependence of the quark
self-energy is directly related to the regularization scheme. In
contrast, in our work, the regularization is independent of the
nonlocality we introduce. It is of interest to note that the value
of the condensate ratio,
$\langle\bar{s}s\rangle/\langle\bar{u}d\rangle$, is about 1.7 when
evaluated using chiral perturbation theory and is only about 1.1
in standard applications of the NJL model. We find that our
nonlocal model can reproduce the larger value of the condensate
ratio when reasonable values are used for the strength of the 't
Hooft interaction. (In an earlier study of the $\eta$(547) and
$\eta$'(958) mesons, we found that use of the larger value of the
condensate ratio led to a very good fit to the mixing angles and
decay constants of these mesons.) We also study the density
dependence of both the quark condensate and the momentum-dependent
quark mass values. Without the addition of new parameters, we
reproduce the density dependence of the condensate given by a
well-known model-independent expression valid for small baryon
density. The generalization of our model to include a model of
confinement required the introduction of an additional parameter.
The further generalization to obtain a nonlocal model also
requires additional parameters. However, we believe our results
are of sufficient interest so as to compensate for the
introduction of the additional parameters in our formalism.
\end{abstract}


\maketitle

\section{INTRODUCTION}

The Nambu--Jona-Lasinio model has been extensively studied for
several decades [1-3]. In recent years there has been strong
interest in the study of quark matter at high densities, using the
NJL model and related models. In particular, one finds color
superconductivity under certain conditions and it has been
suggested that some compact stars might be made of superconducting
quark matter [4-10]. It is our belief that in such studies one
should use a model which reproduces, as well as possible, know
features of QCD. In this work we wish to generalize the
SU(3)-flavor version of the NJL model to be consistent with the
type of momentum-dependent quark masses found in lattice
simulations of QCD [11]. For example, in Figs. 1 and 2 we show
some of the results obtained in Ref. [11]. It may be seen from
these figures that, in Euclidean space, the quark mass goes over
to the current mass for Euclidean momentum $k \gtrsim$ 2 GeV. On
the other hand, the NJL model, in the standard analysis [1], gives
rise to a constant value for the constituent quark mass. As we
will see, the form of the nonlocality used here is different from
that used in Refs. [6, 12-16]. For example, in reference [12] the
$q\bar{q}$ vertex is modified by a form factor which depends on
the relative momentum of the quark and antiquark, while, in
another scheme, a form factor is associated with each quark line
appearing in a diagram. In the latter procedure no further
regularization is needed. However, for the problem considered in
this work, the nonlocal models that appear in the literature are
of limited applicability, since, in the limit of zero current
quark mass, the quark self-energy is proportional to the form
factors used to define the nonlocality [12]. In contrast, in the
current work, we calculate the form of the quark self-energy after
introducing a regulator and a nonlocal quark interaction. We
stress that the procedure used here differs from any that appears
in the literature.

The organization of our work is as follows. In Section II we
review the standard analysis for the condensates and ``gap
equation" of the SU(3)-flavor NJL model [1]. We then go on to
review a procedure for including the contribution to the quark
self-energy due to the addition of a confining interaction for the
case of the SU(2)-flavor model. In Section III we introduce a
nonlocal interaction in the SU(3)-flavor NJL model while
maintaining the separable nature of the interaction. We also
describe the approximation used for the 't Hooft interaction in
the nonlocal model. In Section IV and V we present some of the
results of our numerical calculations. In Section VI we discuss
the density dependence of the quark condensate and the momentum
dependent quark mass. Finally, Section VII contains some
additional discussion and conclusions.

 \begin{figure}
 \includegraphics[
 bb=240 0 250 450,
 angle=-0.5
 , scale=0.5
 ]{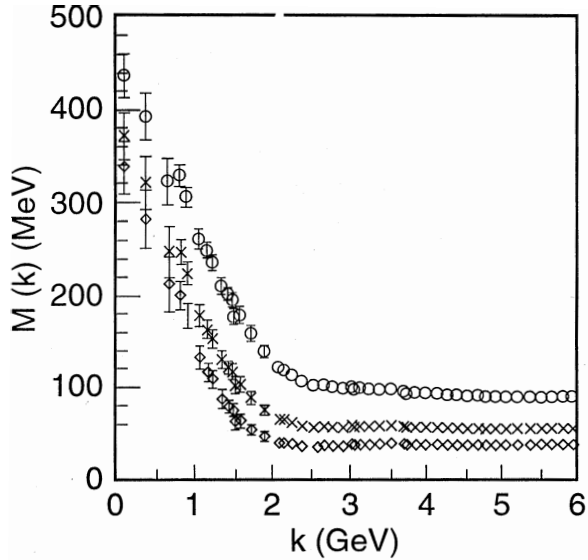}%
 \caption{Quark mass values obtained in Ref. [11]
 for various current quark masses: $m^0=91$ MeV [
 circles], $m^0=54$ MeV [crosses] and $m^0=35$ MeV
 [diamonds].}
 \end{figure}

 \begin{figure}
 \includegraphics[bb=280 20 290 480, scale=0.6]{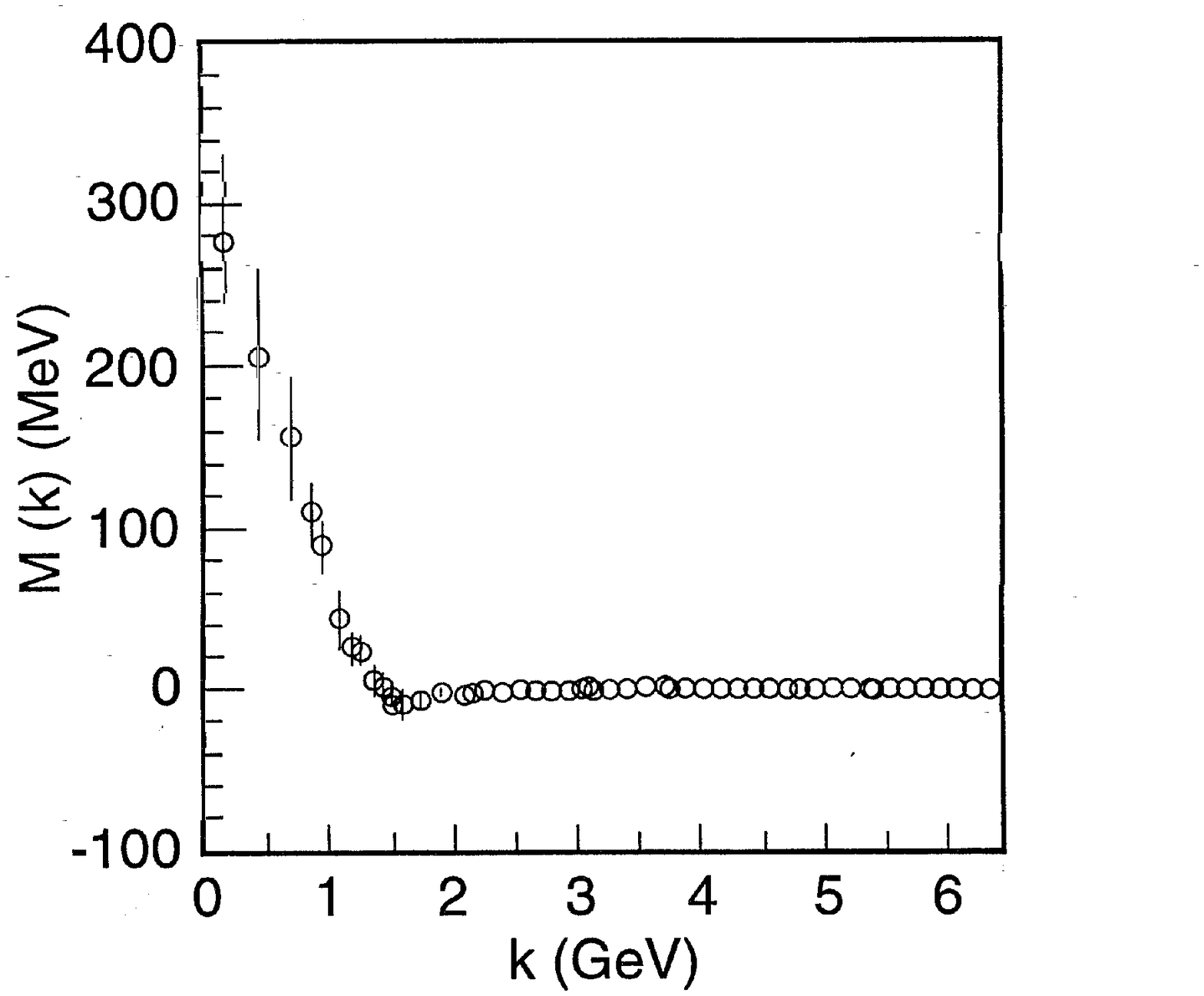}%
 \caption{Quark mass values obtained in Ref. [11]
 using an extrapolation of the current quark mass to zero.
 (The small dip at $k\sim 1.6$ GeV is not statistically significant
 [11].}
 \end{figure}

\section{The quark self-energy in the NJL model}

In order to best introduce the nonlocal model, we will first
review the calculation of the quark self-energy in the local
SU(3)-flavor NJL model and then proceed to add a confinement
interaction, as was done in an earlier study of the quark
self-energy [17]. We consider the generalization to a model with
nonlocal short-range and 't Hooft interactions in the next
section. The Lagrangian of the model is
\begin{eqnarray}
{\cal L}=&&\bar q(i\slr
\partial-m^0)q +\frac{G_S}{2}\sum_{i=0}^8[
(\bar q\lambda^iq)^2+(\bar qi\gamma_5 \lambda^iq)^2]\nonumber\\
&& +\frac{G_D}{2}\{\det[\bar q(1+\gamma_5)q]+\det[\bar
q(1-\gamma_5)q]\}\,.
\end{eqnarray}
Here $m^0$ is the matrix of quark current masses, $m^0 =
\mbox{diag\,}(m_u^0, m_d^0, m_s^0)$ and the $\lambda^i
(i=1,\cdots,8)$ are the Gell-Mann matrices. Further, $\lambda_0 =
\sqrt{2\slash 3}\,\openone$, with \openone being the unit matrix
in the flavor space.

The quark propagator is written as
\begin{equation}
iS(k)=\frac{i}{\slr k-\Sigma(k)+i\epsilon}\,,
\end{equation}
with
\begin{equation}
\Sigma(k)=A(k^2)+B(k^2)\slr k\,.
\end{equation}
We may define
\begin{equation}
M_u(k^2)=\frac{A_u(k^2)}{1-B_u(k^2)}\,,
\end{equation}
and
\begin{equation}
Z_u(k^2)=\frac{1}{1-B_u(k^2)}\,,
\end{equation}
with similar definitions for $M_d(k^2),M_s(k^2),Z_d(k^2)$ and
$Z_s(k^2)$.

In the absence of a confinement model, we have $B(k^2)=0,
A_u(k^2)=A_d(k^2)=m_u$ and $A_s(k^2)=m_s$, where $m_u$ and and
$m_s$ are constants. (Here, we have take $m_u^0=m_d^0$.) In this
case we have [2]
\begin{equation}
m_u=m_u^0-2G_S\langle\bar uu\rangle-G_D\langle\bar
dd\rangle\langle\bar ss\rangle\,,
\end{equation}
\begin{equation}
m_d=m_d^0-2G_S\langle\bar dd\rangle-G_D\langle\bar
uu\rangle\langle\bar ss\rangle\,,
\end{equation}
\begin{equation}
m_s=m_s^0-2G_S\langle\bar ss\rangle-G_D\langle\bar
uu\rangle\langle\bar dd\rangle\,.
\end{equation}
These equations are depicted in Fig. 3a, where the last term
represents the 't Hooft interaction. The up quark vacuum
condensate is given by
\begin{eqnarray}
\langle\bar uu\rangle&=&-N_ci\myint{k}
\texttt{Tr}\frac{C(k^2)}{\slr
k-m_u+i\epsilon}\,,\\\nonumber\\
&=&-4N_ci\myint{k}\frac{m_uC(k^2)}{k^2-m_u^2+i\epsilon}\,.
\end{eqnarray}
Here, $C(k^2)$ is a function needed to regulate the integral. In
this work we will use the Pauli-Villars procedure and evaluate the
integral in Euclidean space, as was done in Ref. [17]. In the
general case we may write
\begin{equation}
\langle\bar uu\rangle=\;-4N_ci\myint{k}
\frac{Z_u(k^2)M_u(k^2)C(k^2)}{k^2-M_u^2(k^2)+i\epsilon}\,,
\end{equation}
where $Z_u(k^2)$ and $M_u^2(k^2)$ were defined in Eqs. (2.4) and
(2.5).

In the appendix of Ref. [17] we considered Lorentz-vector
confinement, with
\begin{equation}
\overline
{V^c}(k_E-k_E^{\,\prime})=\gamma^\mu(1)\gamma_\mu(2)V^c(k_E-k_E^{\,\prime})
\end{equation}
and
\begin{equation}
{V^c}(k_E-k_E^{\,\prime})=\;-8\pi\kappa\left\{\frac{1}{[(k_E-k_E^{\,\prime})^2+\mu^2]^2}
-\frac{4\mu^2}{[(k_E-k_E^{\,\prime})^2+\mu^2]^3}\right\}\,,
\end{equation}
where $k_E^\mu-k_E^{\prime\mu}$ denotes the Euclidean-space
momentum transfer. Here, $\mu$ is a small parameter introduced to
soften the momentum-space singularities. Note that the form
\begin{equation}
{V^c}(\vec k-\vec
k^{\,\prime})=\;-8\pi\kappa\left\{\frac{1}{[(\vec k-\vec
k^{\,\prime})^2+\mu^2]^2} -\frac{4\mu^2}{[(\vec k-\vec
k^{\,\prime})^2+\mu^2]^3}\right\}\,,
\end{equation}
represents the Fourier transform of $V^c(r)=\kappa re^{-\mu r}$,
so that for small $\mu, V^c(r)$ approximates a linear potential
over the relevant range of \emph r. In Fig. 3b we show the
equation for the self-energy when the confining field is included.

In Ref. [17] we obtained the following coupled equations in the
case of the SU(2)-flavor model
\begin{equation}
A(k^2)=\;i\myint{\kp}
\frac{[-4V^c(k-k^{\,\prime})+4N_cn_fG_S]A(\kpsq)}{k^{\,\prime
\,2}[1-B(k^{\,\prime \,2})]^2-A^2(k^{\,\prime \,2})+i\epsilon}\,,
\end{equation}

\begin{equation}
k^2B(k^2)=\;i\myint{\kp}\frac{2(k\cdot
k^{\,\prime})[1-B(k^{\,\prime
\,2})]V^c(k-k^{\,\prime})}{k^{\,\prime \,2}[1-B(k^{\,\prime
\,2})]^2-A^2(k^{\,\prime \,2})+i\epsilon}\,.
\end{equation}
These equations were solved after passing to Euclidean space and
including a Pauli-Villars regulator of the form
\begin{equation}
C(k_E^{\,2})=
\frac{2\Lambda^4}{[k_E^{\,2}+A^2(k_E^{\,2})+\Lambda^2][k_E^{\,2}+A^2(k_E^{\,2})+2\Lambda^2]}
\end{equation}
in Euclidean space. Note that the form
\begin{equation}
\widetilde C(k_E^{\,2})=
\frac{2\Lambda^4}{[k_E^{\,2}\left(1-B(k_E^{\,2})\right)^2
+A^2(k_E^{\,2})+\Lambda^2][k_E^{\,2}\left(1-B(k_E^{\,2})
\right)^2+A^2(k_E^{\,2})+2\Lambda^2]}
\end{equation}
may also be used. (In Eqs. (2.15) and (2.16) $V^c$ appears with a
sign opposite to that given in Ref. [17], since, in that work, we
used a negative value of $\kappa$. For the present work we use a
positive value of $\kappa$ to be consistent with all of our other
publications.)

 \begin{figure}
 \includegraphics[bb=180 25 190 250, angle=-0.2, scale=0.8]{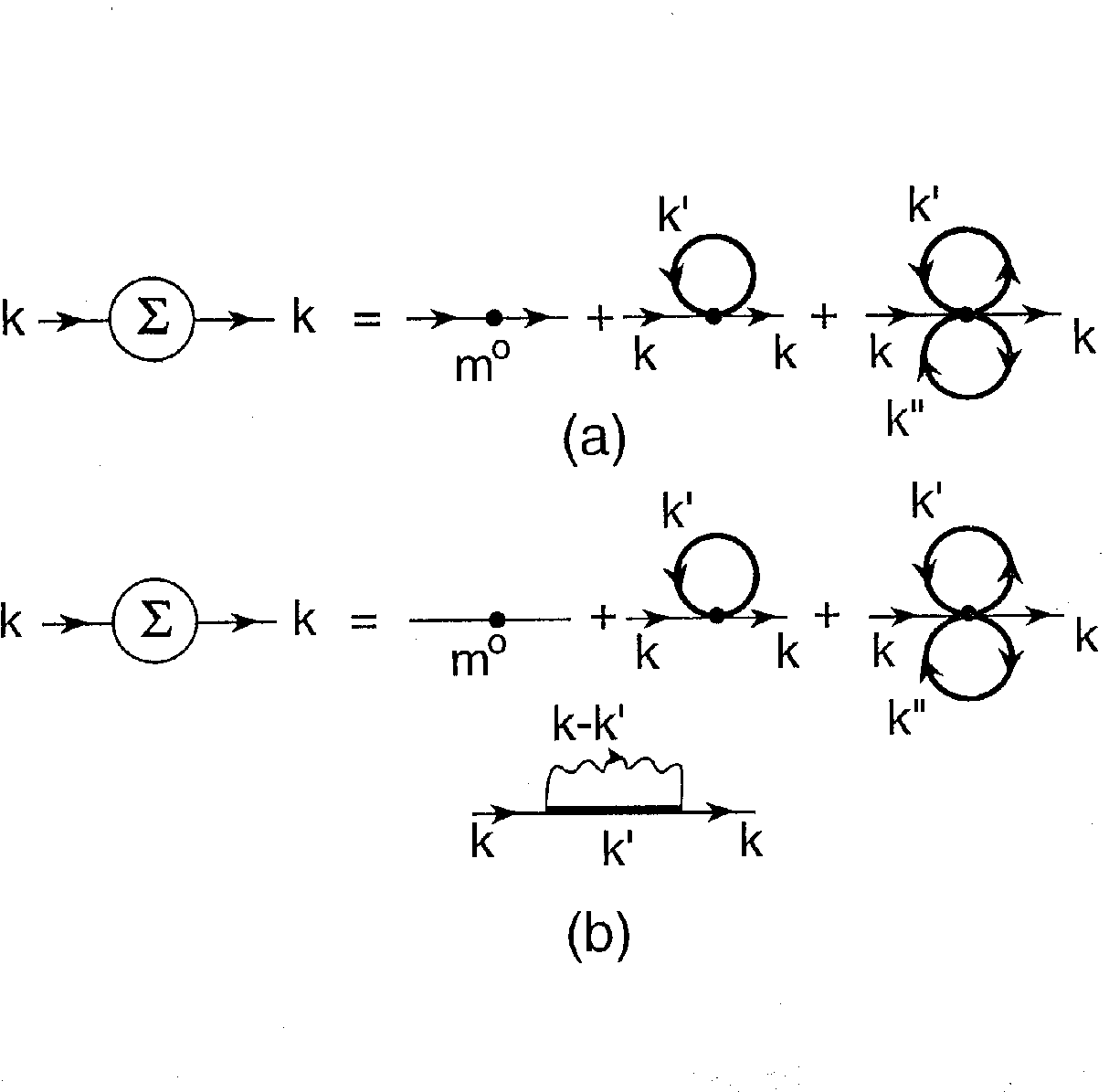}%
 \caption{a)A diagrammatic representation of the equation
 for the self-energy for a quark of momentum $k$. The first
 term on the right is the contribution of the current quark mass $m^0$.
 The second term corresponds to the term proportional to $G_S$ in Eqs. (3.6)--(3.8)
 and the last term represents the 't Hooft interaction.
 b)The self-energy equation in the presence of a confining interaction (wavy line.)
 Without the 't Hooft interaction, we have a representation of SU(2)-flavor
 model studied in Ref. [17]. (See Eqs. (2.15) and (2.16).)}
 \end{figure}

\section{A nonlocal NJL model}

In this section we describe the procedures we use to create a
nonlocal version of our generalized NJL model. We consider the
second term of Eq. (2.1) and make the replacement
\begin{eqnarray}
\frac{G_S}{2}&&\sum_{i=0}^8[(\bar q(x)\lambda_iq(x))^2+(\bar
q(x)i\gamma_5\lambda_iq(x))^2]\\\nonumber
&&\longrightarrow\frac{G_S}{2}\sum_{i=0}^8\{[\bar
q(x)\lambda^if(x)q(x)\cdot \bar q(y)\lambda^if(y)q(y)\\[0.4cm]\nonumber
&&\hspace{0.9cm}+[\bar q(x)i\gamma_5\lambda^if(x)q(x)\cdot\bar
q(y)i\gamma_5\lambda^if(y)q(y)]\}\,.
\end{eqnarray}
This replacement corresponds to the use of a separable interaction
$V(x-y)=G_Sf(x)f(y)$.

A related modification may be made for the 't Hooft interaction.
It is useful, however, to describe these modifications as they
affect momentum-space calculations. With reference to Fig. 4, we
replace $G_S$ by $f(k_1-k_2)G_Sf(k_3-k_4).$ In the evaluation of
the second term of Fig. 2 we need
$G_S(k-k^{\,\prime})=f(k-k^{\,\prime})G_Sf(k-k^{\,\prime})$ and
choose to write
\begin{eqnarray}
G_S(k-k^{\,\prime})&=&\:\exp[-(k-k^{\,\prime})^{\,2n}/2\beta]
G_S\exp[-(k-k^{\,\prime})^{\,2n}/2\beta]\\\nonumber
&=&\:G_S\exp[-(k-k^{\,\prime})^{\,2n}/\beta]\,.
\end{eqnarray}
In this work we take $n=4$ and $\beta=20$ GeV${}^{8}$. In Fig. 5
we exhibit the function $F(k^2)=\exp[-k^{\,2n}/\beta]$. It is
clear that many other functions may be chosen.

We now rewrite Eqs. (2.15) and (2.16) for the up, down and strange
quarks. For example, for the SU(3)-flavor case
\begin{eqnarray}
A_u(k^2)&=&m_u^0+i\myint{\kp}
\frac{[-4V^c(k-k^{\,\prime})+8N_cG_S(k-k^{\,\prime})]A_u(k^{\,\prime\,2})}
{k^{\,\prime\,2}[1-B_u(k^{\,\prime\,2})]^2-A_u^2(k^{\,\prime\,2})+i\epsilon}\,,
\\[0.35cm]
k^2B_u(k^2)&=&i\myint{\kp}
\frac{2(k\cdot\kp)[1-B_u(\kpsq)]V^c(k-\kp)}
{k^{\,\prime\,2}[1-B_u(k^{\,\prime\,2})]^2-A_u^2(k^{\,\prime\,2})+i\epsilon}\,,
\end{eqnarray}
with similar equations for $A_d(k^2), B_d(k^2),$ etc. Again, these
equations are solved after passing to Euclidean space and
introducing regulator functions: $C_u(\ksq),\, C_d(\ksq) \texttt{
and }C_s(\ksq)$. In Euclidean space we write \be
C_u(\ksq)=\frac{2\Lambda^4}{[\ksq+A_u^2(\ksq)+\Lambda^2][\ksq+A_u^2(\ksq)+2\Lambda^2]}\,,
\ee etc. Note that without the 't Hooft interaction the equations
for the up, down and strange quarks are uncoupled.

 \begin{figure}
 \includegraphics[bb=280 75 290 250, scale=0.6]{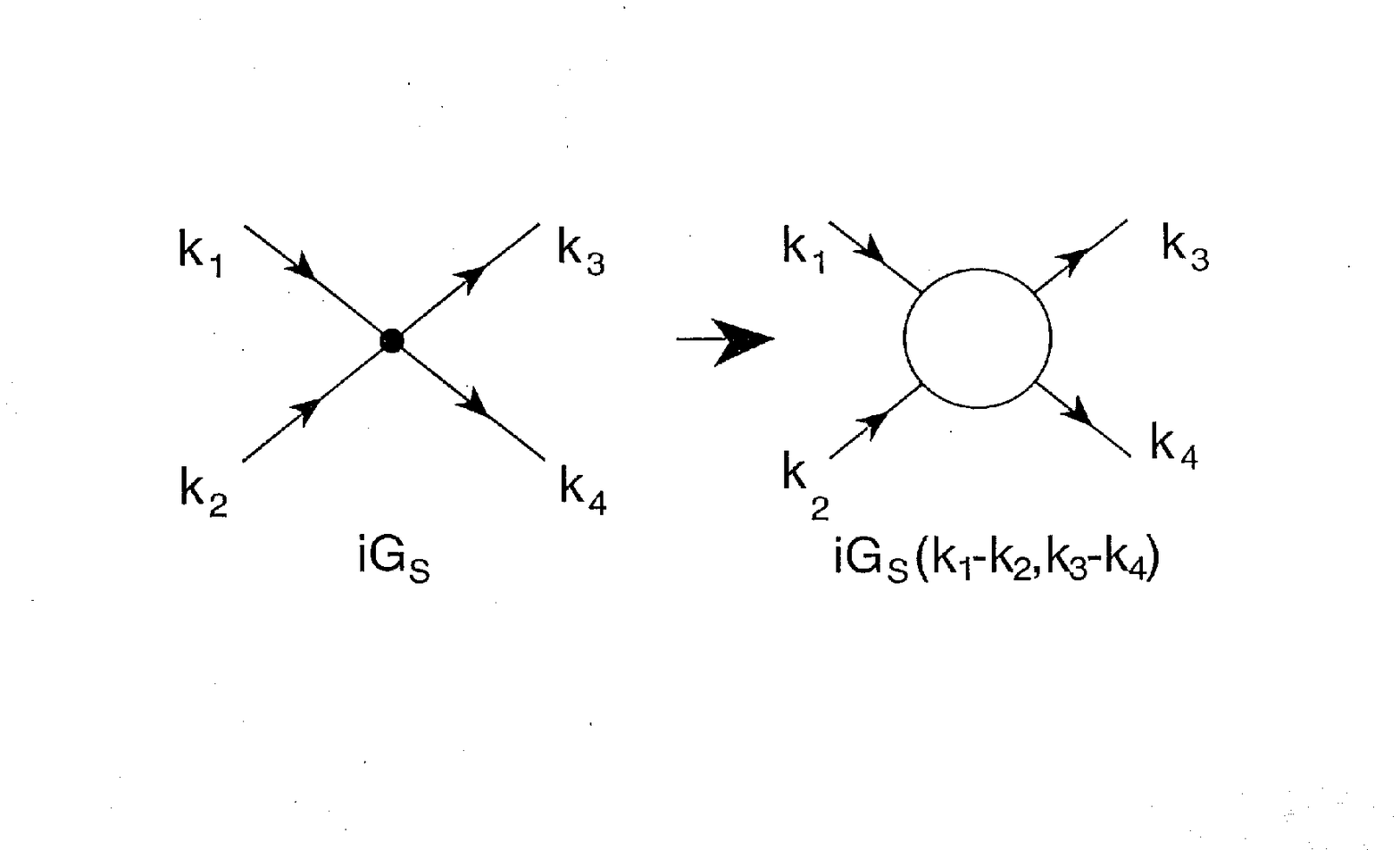}%
 \caption{The figure indicates the replacement of the local quark interaction, $iG_S$,
 by the nonlocal (separable) term, $iG_S(k_1-k_2, k_3-k_4)=iG_Sf(k_1-k_2)f(k_3-k_4)$.
 The distinction between our separable model and that of Ref. [12], for example, is that
 in Ref. [12] the alternative replacement $iG_S\longrightarrow iG_S(k_2+k_4, k_1+k_3)
 =iG_Sf(k_2+k_4)f(k_1+k_3)$ was used.}
 \end{figure}

 \begin{figure}
 \includegraphics{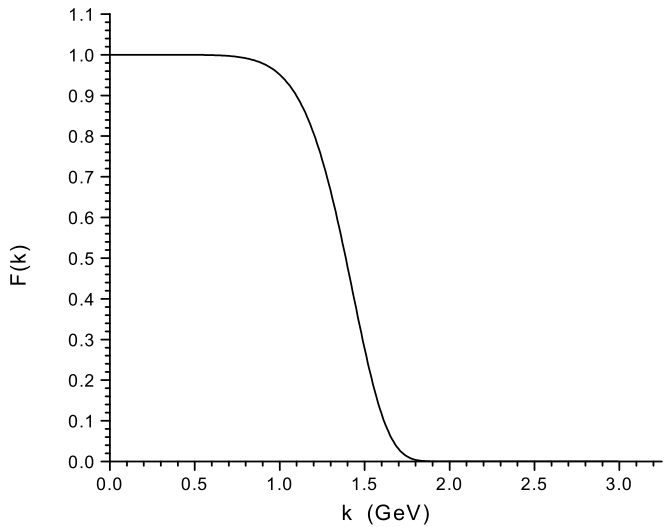}%
 \caption{The correlation function $F(k)=\exp[-k^{\,2n}/\beta]$ is
 shown for $n=4$ and $\beta=20$\gev{8}.}
 \end{figure}

Our treatment of the 't Hooft interaction is based upon a
generalization of the last term in Eqs. (2.6)-(2.8). With
reference to the third term on the right in Fig. 6, we introduce a
correlation between the quark of momentum \emph{k} and the quark
of momentum $\kp$. We also include a correlation between the quark
of momentum \emph{k} and that of momentum $\kp{}^\prime$.
(Therefore, our procedure does not introduce a correlation between
the two quarks in the separate condensates. At this stage of the
development of our model that seems to be a reasonable
approximation and avoids having to define a three-quark
correlation function, $f(k-\kp, \kp-\kdp, k-\kdp).$) For example,
we generalize the term $-G_D\langle\bar dd\rangle\langle\bar
ss\rangle$ to obtain the contribution to $A_u(k)$:\be A_u^t(k)=
-G_D\left[-4N_ci\myint{\kp}\frac{C_d(\kpsq)Z_u(\kpsq)M_u(\kpsq)f^2(k-\kp)}
{\kpsq[1-B_u(\kpsq)]^2-A_u^2(\kpsq)}\right]
\\[0.35cm]\nonumber \hspace{0.6cm}\times\left[-4N_ci\myint{\kdp}
\frac{C_s(\kdp{}^2)Z_s(\kdp{}^2)M_s(\kdp{}^2)f^2(k-\kdp)}
{\kdp{}^2[1-B_s(\kdp{}^2)]^2-A_s^2(\kdp{}^2)}\right] \ee in
Minkowski space. This expression is then evaluated in Euclidean
space and added to the right-hand side of Eq. (3.3). In a similar
fashion, we calculate $A_d^t(k)$ and $A_s^t(k)$.

 \begin{figure}
 \begin{center}
 \includegraphics[bb=200 50 210 250, angle=-2, scale=0.8]{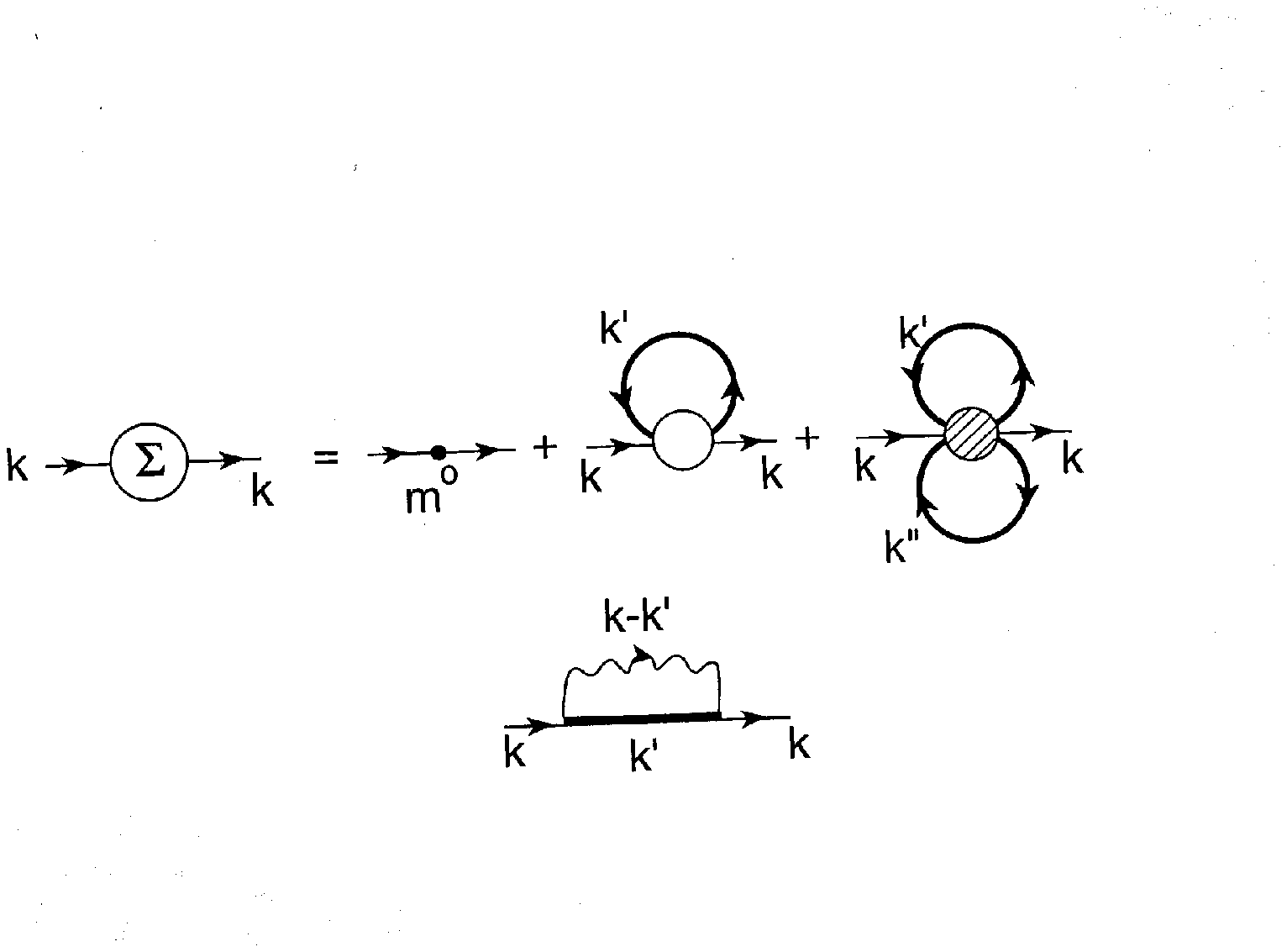}%
 \caption{The quark self-energy equation is depicted, with nonlocal terms replacing
 $G_S$ and $G_D$. [See Fig. 4.]}
 \end{center}
 \end{figure}

\section{Numerical Results: Condensates and Constituent Mass Values}

There is a good deal of flexibility in choosing the regulators
$C_u(\ksq)$, $C_d(\ksq)$, and $C_s(\ksq)$. Also, various forms
could be chosen for the correlation functions, $f(k-\kp)$.
Previously, in our Minkowski-space studies of the $\eta$ mesons we
used $G_S = 11.84$ GeV${}^{-2}$ and $G_D\simeq -200$\gev{-5} [18].
However, in that work we used a Gaussian regulator in Minkowski
space so that a direct comparison with the present study can not
be made. On the other hand, we do not expect to find radically
different parameters, if the constituent masses in the two
calculations are similar. For example, in our earlier work, in
which $m_u$ and $m_s$ were parameters, we used $m_u=0.364$\gev{},
which can be compared to the value of $M_u(0)$ calculated here. We
have also used either $m_s=0.565$\gev{} [19-23] or
$m_s=0.585$\gev{} [18], values which may be compared to $M_s(0)$.

To proceed, we take $\Lambda=1.0$\gev{}, $G_S=13.30$\gev{-2},
$\kappa=0.055$\gev{2}, $m_u^0=0.0055$\gev{}, $m_s^0=0.130$\gev{},
$\mu=0.010$\gev{} and $\beta=20.0$\gev{8}. We then consider
values of $G_D=0$, $G_D=-20G_S$, $G_D=-30G_S$ and $G_D=-40G_S$.
The results of our calculations are given in Table I. Recall that
the function $F(k)$ does not appear in our expression for the
condensates. The calculation of the condensates includes the
Pauli-Villars regulators, $C_u(\ksq)$, $C_d(\ksq)$ and
$C_s(\ksq)$, however. [See Eqs. (2.9)-(2.11).] In our calculation
of the properties of the $\eta$ mesons [18] we had $-G_D/G_S\simeq
15-18$, since we used $G_D$ values in the range
$-180$\gev{-5}$\leq G_S\leq -220$\gev{-5} in that work.

In order to specify a value of $G_D$ for this work, we note that a
calculation based upon chiral perturbation theory yields
$\langle\bar ss\rangle/\langle\bar uu\rangle = 1.689$ [24].
Inspection of Table I suggests that the values of $G_D$, other
than $G_D=0$, given in Table I are acceptable. For
$G_D=-266$\gev{-5} we have $M_u(0)= 0.377$\gev{} and $M_s(0)=
0.555$\gev{}, which are reasonably close to the phenomenological
parameters $m_u= 0.364$\gev{} and $m_s= 0.565$\gev{} used in our
earlier work [19-23].

It is worth noting that, in standard application of the
SU(3)-flavor NJL model, one finds $\langle\bar
ss\rangle/\langle\bar uu\rangle \sim 1.1$ [1], so that the results
shown in Table I are encouraging, given that the value for
$\langle\bar ss\rangle/\langle\bar uu\rangle$ obtained using
chiral perturbation theory is about 1.7 [24], as noted above.

\begin{table}
 \begin{tabular}{||@{\hspace{0.2cm}}l@{\hspace{0.2cm}}||@{\hspace{0.5cm}}
 l@{\hspace{0.5cm}}|@{\hspace{0.5cm}}l@{\hspace{0.5cm}}
 |@{\hspace{0.5cm}}l@{\hspace{0.5cm}}|@{\hspace{0.5cm}}l@{\hspace{0.2cm}}||}\hline\hline
 $G_D$ [\gev{-5}]    &0.0     &-266.0 &-399.0 &-532.0\\\hline
 $M_u(0)$ [GeV]      &0.334   &0.377  &0.396  &0.416\\\hline
 $M_s(0)$ [GeV]      &0.538   &0.555  &0.564  &0.575\\\hline
 $\langle\bar uu\rangle^\frac{1}{3}$ [GeV]
                      &-0.207  &-0.215 &-0.217 &-0.220\\\hline
 $\langle\bar ss\rangle^\frac{1}{3}$ [GeV]
                      &-0.2605 &-0.261 &-0.261 &-0.261\\\hline
 $\frac{\langle\bar ss\rangle}{\langle\bar uu\rangle}$
                      &2.00    &1.80   &1.73   &1.68  \\\hline
 $A_u(0)$ [GeV]      &0.447   &0.481  &0.496  &0.512 \\\hline
 $B_u(0)$            &-0.335  &-0.276 &-0.253 &-0.233\\\hline
 $A_s(0)$ [GeV]      &0.614   &0.628  &0.636  &0.645  \\\hline
 $B_s(0)$            &-0.139  &-0.131 &-0.127 &-0.122\\\hline\hline
 \end{tabular}
 \vspace{1.2cm}
 \caption{Calculated values for the condensates and for $A(0),
B(0), \mbox{and } M(0)$ are given for the up and strange quarks
for four values of $G_D$. The parameters $m_u^0=0.0055$ GeV,
$m_s^0=0.130$ GeV, $\kappa=0.055$\gev2, $\beta=20$\gev8,
$\mu=0.010$ GeV, $\Lambda=1.0$ GeV, $G_S=13.30$\gev{-2} were used.
The values of $\kappa$ and $\mu$ were fixed in earlier work
[18-23]. Values of $\langle\bar uu\rangle\simeq\langle\bar
dd\rangle\simeq -(0.240\pm 0.025\mbox{ GeV}){}^3$ have been
suggested [28], so we see that our calculated values are at, or
near, the lower limit for that quantity.}
 \end{table}

\section{numerical results:\\momentum dependence of the constituent quark masses}

In Fig. 7 we show $M_u(k)$, where \emph k is the magnitude of the
Euclidean momentum. The dashed line exhibits the result without
the confining interaction ($\kappa=0$). It is interesting to see
that inclusion of confinement improves the shape of the curve when
we compare our results to the lattice results shown in Figs. 1 and
2. We note that $M_u(k)$ goes over to $m_u^0=0.0055$\gev{} for
large \emph k. In Fig. 8 we show $B_u(k)$. (Recall that
$Z_u(k)=[1-B_u(k)]^{-1}$.) We remark that $B_u(k)=0$ when
$\kappa=0$. In Figs. 9 and 10 we show $M_s(k)$ and $B_s(k)$,
respectively. As expected, we find that $M_s(k)$ goes over to
$m_s^0=0.130$\gev{} when \emph k is large.

 \begin{figure}
 \includegraphics{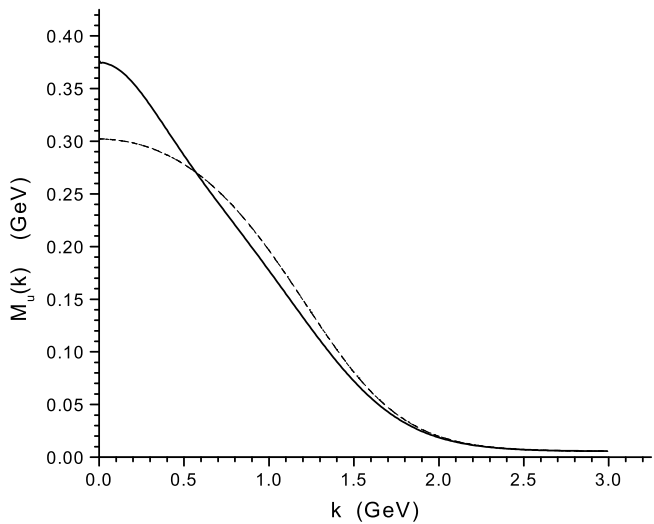}%
 \caption{Values of $M_u(k)$ are shown for the parameters $m_u^0=0.0055$ GeV,
 $m_s^0=0.130$ GeV, $\kappa=0.055$\gev{2}, $\mu=0.010$ GeV, $\Lambda=1.0$ GeV,
 $\beta=20$\gev{8}, $G_S=13.3$\gev{-2}, and $G_D=-266$\gev{-5}. The dashed line
 shows the result without confinement ($\kappa=0$).}
 \end{figure}

 \begin{figure}
 \includegraphics{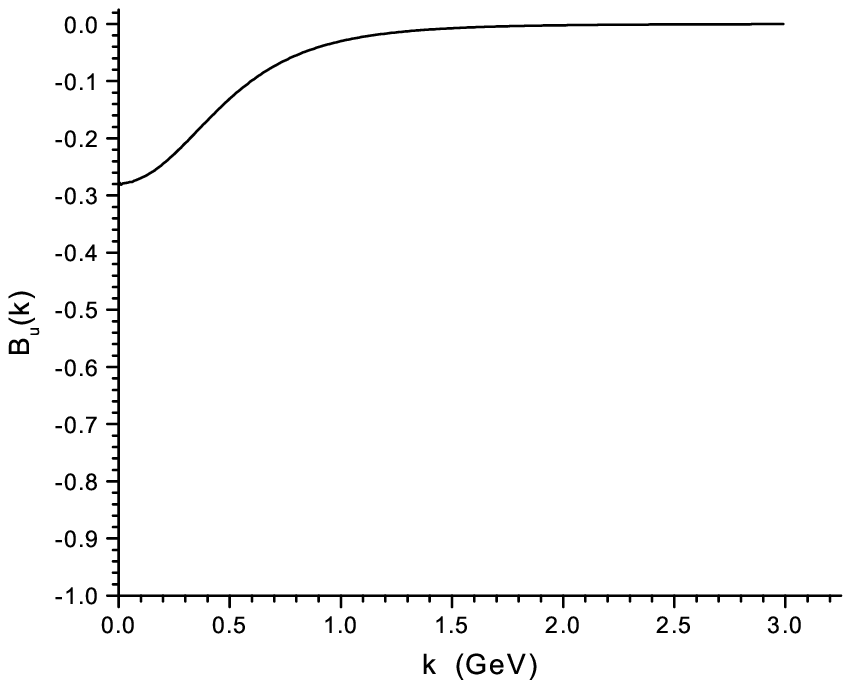}%
 \caption{Values of $B_u(k)$ are shown. (See caption to Fig. 7.)}
 \end{figure}

 \begin{figure}
 \includegraphics{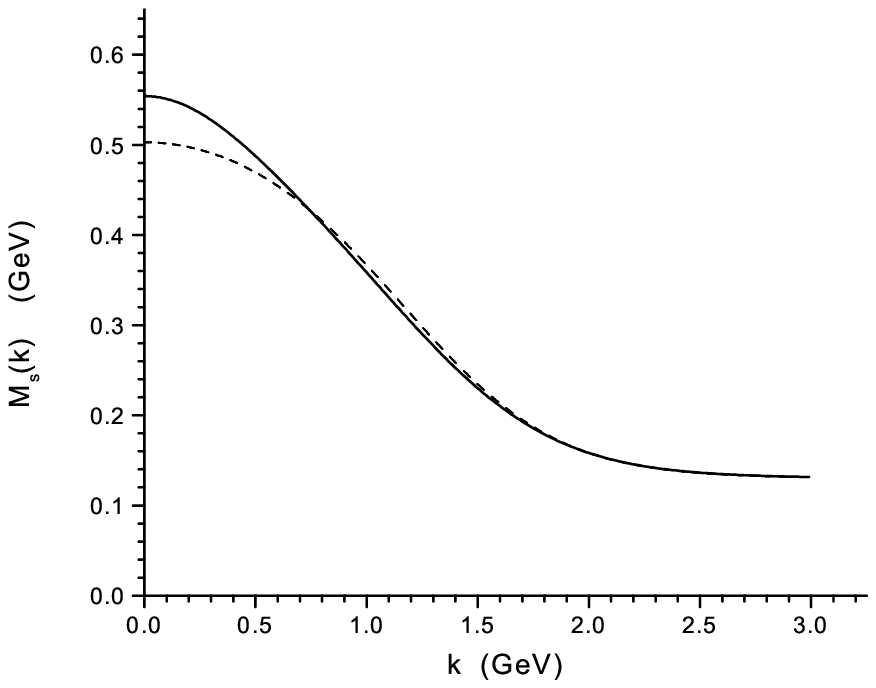}%
 \caption{Values of $M_s(k)$ are shown. (See caption to Fig. 7.)}
 \end{figure}

 \begin{figure}
 \includegraphics{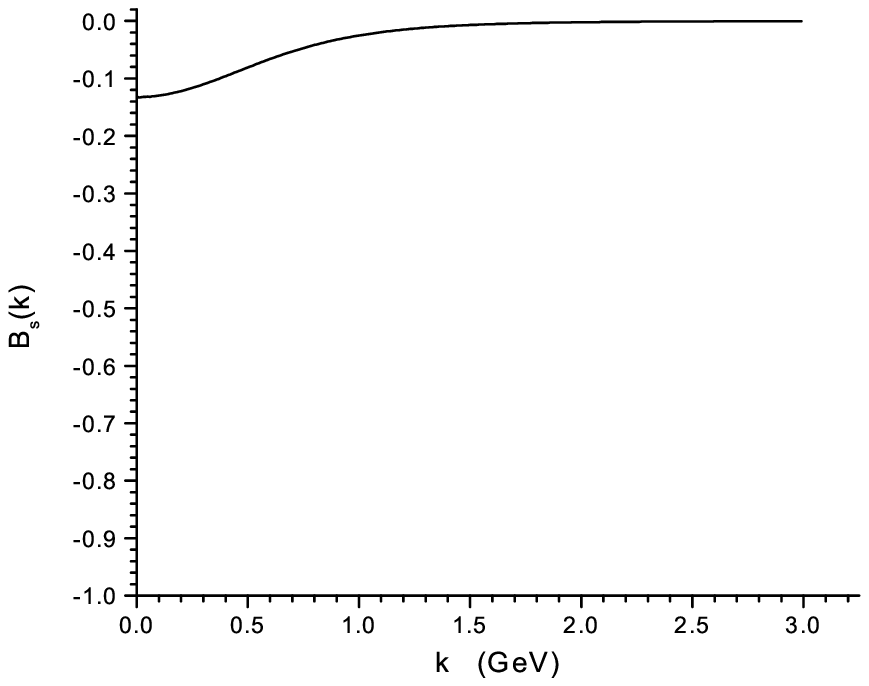}%
 \caption{Values of $B_s(k)$ are shown. (See caption to Fig. 7.)}
 \end{figure}

\section{density dependence of the quark mass and quark
condensate}

As stated earlier, the behavior of the NJL model for finite values
of the baryon density is an extensively explored topic [1, 25,
26], with particular recent emphasis on color superconductivity
[4-10]. In this section we explore the behavior of our model at
finite baryon density. (It should be noted that the nature of the
phase transition describing chiral symmetry restoration at finite
density is quite model dependent. For example, the inclusion of
current quark masses can change a strong first-order transition to
a smooth second-order transition [26].)

A comprehensive study of the thermodynamics of the three-flavor
NJL model has been reported in Ref. [27]. There it is found that
the up, down and strange quark masses are essentially constant up
to the density where a first-order phase transition appears. At
that point, the up and down quark masses drop from a value of
about 380 MeV to about 30 MeV. That behavior differs from the
behavior expected at low density. For example, we have the
well-known relation between the value of the condensate and the
baryon density of nuclear matter \be \frac{\langle\bar
qq\rangle_\rho}{\langle\bar
qq\rangle{}_0}=1-\frac{\sigma_N\rho_B}{f_\pi^2m_\pi^2}\,, \ee
where $\sigma_N$ is the pion-nucleon sigma term. This relation is
valid to first-order in the density. It may be derived, in the
case of nuclear matter, by writing \be \langle\bar qq\rangle_\rho
= \langle\bar qq\rangle_0+\langle N|\bar qq|N\rangle \rho_B \ee
and making use of the definition of the pion-nucleon sigma term,
$\sigma_N$, and the Gell-Mann--Oakes--Renner relation. If we put
$\sigma_N=0.045$ GeV, we have $\langle\bar
qq\rangle_\rho/\langle\bar qq\rangle_0=1-0.273\rho_B$, where
$\rho_B$ is in\gev{3} units. For nuclear matter $\rho_B=$(\,0.109
GeV)${}^3$, so we see that the condensate is reduced by about
35\%, if we evaluate Eq. (6.1) at nuclear matter density. We can
check whether the density dependence given by Eq. (6.1) is
reproduced in our model, since it should not matter whether the
scalar density of the background matter is generated by quarks in
nucleons or by the presence of free quarks. In the former case, we
may write, for the baryon density, \be
\rho_B=4\!\int^{\,k_F}\hspace{-0.3cm}\frac{d^3\!k}{(2\pi)^3} \ee
where the factor of 4 is arises from the product of the spin and
isospin factors. In the case of quarks, we have \be
\rho_B=4N_c\left(\frac{1}{3}\right)
\!\int^{\,k_F}\hspace{-0.3cm}\frac{d^3\!k}{(2\pi)^3} \ee where, in
this case, the factor of 4 again arises from the spin and isospin
factor. (Both up and down quarks are present in equal numbers.)
The color factor, $N_c=3$, is cancelled by the baryon number of
1/3 of each quark.

We need to modify the equations for the quark self-energy to take
into account the presence of the Fermi seas of up and down quarks
whose Fermi momentum is $k_F$. We take one Fermi sea to be
composed of on-mass-shell up quarks with constituent mass
$M_u(0)$. The following term is then added to the equation for
$A_u(k)$. \be A_u^{(\rho)}(k)=-(2G_S)N_c2\!\int^{\,k_F}
\hspace{-0.3cm}\frac{d^3\!\kp}{(2\pi)^3}\,\frac{M_u(0)}{E_u(k)}f^2(k-\kp)
\ee where $E_u(k)=\left[\vec k^2+M_u^2(0)\right]^\frac{1}{2}$. The
second factor of 2 in Eq. (6.5) reflects the spin degeneracy. We
note that $M_u(0)$ is density-dependent and could be written as
$M_u(0,\rho)$ in keeping with the labelling of Fig. 11, where
$M_u(k,\rho)$ is used. Also, if $f(k-\kp)=1$, $A_u^{(\rho)}(k)$
would then represent $-2G_S\,\rho_S$, where $\rho_S$ is the scalar
density associated with the up-quark Fermi sea.

In Fig. 11 we show $M(k,\rho)$ calculated for four values of
$\rho$ and in Fig. 12 we show \emph M(0) as a function of $k_F^3$.
Since the quarks in the Fermi sea are taken to be on-mass-shell,
we would, in principle, require $M_u(k,\rho)$ in Minkowski space.
However, since $k_F$ = 0.268 GeV for the case of nuclear matter,
only a very modest extrapolation of the curves shown in Fig. 11 is
needed for the densities considered in this work. In Fig. 13 we
show the value of the up quark condensate as a function of
$k_F^3$. (Note that $k_F^3 = 19.2\times 10^{-3}$\gev{3} represents
the density of nuclear matter.) It is seen that, for small values
of the density, the density dependence of the condensate
reproduces what is expected from Eq. (6.1). (If we extrapolate the
curve using a linear approximation, the condensate is reduced by
about 30\% at nuclear matter density.)

 \begin{figure}
 \includegraphics{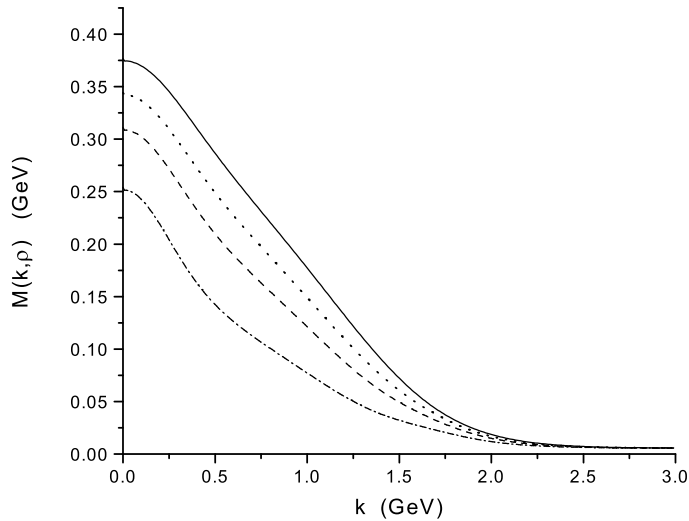}%
 \caption{The values of $M(k,\rho)$ are shown for $\rho/\rho_{NM}=0.26$ [dotted line],
 $\rho/\rho_{NM}=0.52$ [dashed line], and $\rho/\rho_{NM}=0.78$ [dash-dot line].}
 \end{figure}

 \begin{figure}
 \includegraphics{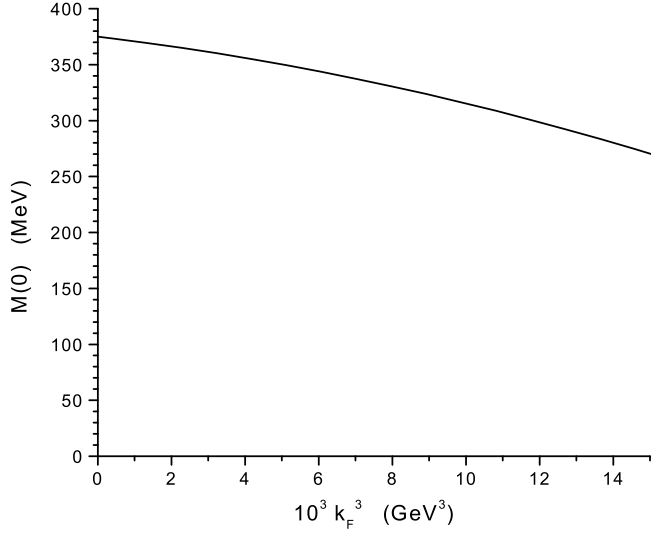}%
 \caption{Values of $M(0)$ are shown as a function of $k_F^3$. Note that $\rho_B
 =(2/3\pi^2)k_F^3.$}
 \end{figure}

 \begin{figure}
 \includegraphics{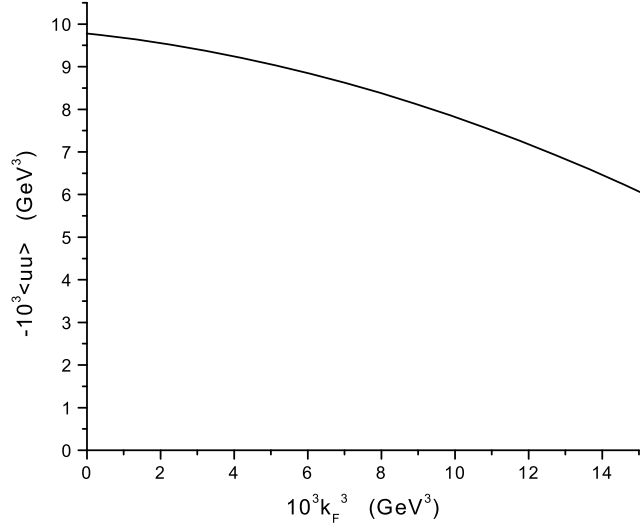}%
 \caption{The values of the up quark condensate are
 given as a function of $k_F^3$. Note that $\rho_B
 =(2/3\pi^2)k_F^3.$}
 \end{figure}

\section{discussion}

We have remarked earlier in this work that the large values of the
condensate ratio $\langle\bar ss\rangle/\langle\bar uu\rangle$
seen in Table I play a role in obtaining a good fit to the mixing
angles of the $\eta$(947) and $\eta\prime$(958) mesons [18]. To
understand this remark we note that the effective singlet-octet
coupling constants for pseudoscalar states are [3] \be
G_{00}^P&=&G_S-\frac{2}{3}(\alpha+\beta+\gamma)\frac{G_D}{2}\,,\\
G_{88}^P&=&G_S-\frac{1}{3}(\gamma-2\alpha-2\beta)\frac{G_D}{2}\,,
\ee and \be
\hspace{-0.5cm}G_{08}^P=-\frac{\sqrt2}{6}(2\gamma-\alpha-\beta)\frac{G_D}{2}\,,
\ee where $\alpha=\langle\bar uu\rangle$, $\beta=\langle\bar
dd\rangle$ and $\gamma=\langle\bar ss\rangle$. We take
$\alpha=\beta$, so that \be
G_{08}^P=-\frac{\sqrt2}{3}(\gamma-\alpha)\frac{G_D}{2}\,.\ee If
$\gamma=1.7\alpha$, the result for $G_{08}^P$ is six times larger
than when $\gamma=1.1\alpha$.

In addition to the effects of $G_{08}^P$, singlet-octet mixing is
induced by the quantity [18] \be
E_{08}(k)=\frac{2\sqrt2}{3}[E_u(k)-E_s(k)]\,,\ee where
$E_u(k)=\left[\vec k^2+m_u^2\right]^\frac{1}{2}$, etc. It is found
that, since $G_{08}^P$ and $E_{08}(k)$ tend to cancel in our
formalism, the significant singlet-octet mixing generated by
$E_{08}(k)$ is reduced by the values of $G_{08}^P$ obtained for
the larger value of the ratio $\langle\bar ss\rangle/\langle\bar
uu\rangle$, with the result that we reproduce the values of the
mixing angles found in other studies that make use of experimental
data to obtain values for the mixing angles [18].

In our earlier work, which was carried out in Minkowski space, the
values of $m_u=m_d$ and $m_s$ were taken as parameters. Inspection
of our figures which exhibit values of $M_u(k)$ and $M_s(k)$
suggests that an extrapolation into Minkowski space may be made if
$\ksq$ is not too large. The fact that $M_u(0)$ and $M_s(0)$ are
close to our phenomenological parameters for $G_D=-266$\gev{-5} is
encouraging and suggests that some support for our choice of quark
mass parameters may be found in our Euclidean-space analysis.

The full consequences of separating the specification of the
nonlocality of the quark interaction from the choice of the
regulator of the theory should be explored more fully. Although
that feature of our model introduces greater flexibility, that
comes with the disadvantage of having to introduce other
parameters in the model. We have made only limited variation of
the form of the nonlocality and the regulator. For further
applications it may be of interest to explore a more comprehensive
parameter variation. It is also necessary to extend the
calculations reported in Figs. 11--13 to larger values of the
density than those considered here. That step will require more
complex methods for solving our nonlinear equations for the
self-energy than the simple iteration scheme we have used thus
far.

Our work may be compared to that of Alkofer, Watson and Weigel
[29] who have solved the Schwinger-Dyson equation using a gluon
propagator whose low-momentum behavior is enhanced by a Gaussian
function. (That modification requires the introduction of two
phenomenological parameters [30].) The behavior found for $A(k)$
and $B(k)$ in Euclidean space is similar to that obtained in this
work. (See Fig. 1 of Ref. [29].) Those authors also solve the
Bethe-Salpeter equation to obtain the properties of various $q\bar
q$ mesons with generally satisfactory results. It is of interest
to note that the Minkowski space solution for $A(k)$ and $B(k)$ is
such that the quark can go on-mass-shell. That feature may be
related to our work [18-23] in which we use on-mass-shell quarks
with masses $m_u=m_d=0.364$ GeV and $m_s=0.565$ GeV (or 0.585 GeV
[18]) when solving the Bethe-Salpeter equation in our study of
$q\bar q$ mesons.


\vspace{1.5cm}
\noindent$\textbf{References}$\\[-2cm]


\end{document}